\documentclass[twocolumn,showpacs,preprintnumbers]{revtex4}
\usepackage{amsmath}
\usepackage{amssymb}
\usepackage{graphicx}
\usepackage{dcolumn}
\usepackage{bm}
\usepackage{amssymb}
\usepackage{graphicx}
\usepackage{dcolumn}
\usepackage{bm}
\usepackage{soul,xcolor}
\begin{document}
\graphicspath{{./img/}}
\definecolor{zbgreen}{rgb}{0,0.6,0}
\setstcolor{red}   
\title{Symmetry breaking and entropy production during the evolution of spinor Bose-Einstein condensate driven by coherent atom beam }

\author{Yixin Xu$^{1}$, Zhongda Zeng$^{1}$, Zbigniew Domanski$^{2}$, and Zhibing Li$^{1,3,4}$\footnote{Corresponding author: Z.B.Li, stslzb@mail.sysu.edu.cn}}

\affiliation{$^{1}$School of Physics, Sun Yat-Sen University, Guangzhou, 510275, P. R. China}
\affiliation{$^{2}$Institute of Mathematics, Czestochowa University of Technology, 42-201 Czestochowa, Poland}
\affiliation{$^{3}$State Key Laboratory of Optoelectronic Materials and Technologies, Guangzhou, 510275, P. R. China}
\affiliation{$^{4}$Guangdong Province Key Laboratory of Display Material and Technology, Guangzhou, 510275, P. R. China}

\begin{abstract}
The spinor condensate with spin states degenerated in the ground spin-space provides a unique platform for investigating the edge of quantum mechanics and statistical physics. We study the evolution of the condensate under the scattering of a coherent atom beam. The time-dependent magnetization, entanglement entropy, thermal entropy, and the entropy production rate are calculated. A novel spontaneous symmetry breaking is found during the evolution. It is shown that the stationary spin distribution can be controlled by the incoming coherent spin state of the incident atom beam, therefore the atom-condensate scattering provides a new way to probe the spin distribution of the condensate.
\end{abstract}

\maketitle
Statistical behavior of quantum systems with few degrees of freedom would exhibit subtle interplay between quantum statistics and quantum mechanics.  In recent years, a number of ingenious experiments on the thermalization of small quantum systems have been carried out.\cite{eise15,isla15,kauf16,wei18} They revealed that the quantum entanglement is indispensable for validating the eigenstate thermalization hypothesis of equivalence between the thermal ensemble averages and the quantum-mechanical time averages of observable quantities in isolated quantum systems that were initialized in coherent superpositions of eigenstates.\cite{deut91,sred94,rigo08}
Recently the entropy production rate at the steady state of a Bose-Einstein condensate (BEC) coupled with a micromechanical resonator has been measured and irreversibility of the system has been discussed.\cite{brun18b}
On the other hand, the BEC is an example of spontaneous symmetry breaking (SSB), where the gauge symmetry is broken below the critical temperature. For a BEC having spin degrees of freedom, there would be another type of SSB in the degenerated space of states with the lowest energy. Any new mechanism for SSB is of broad interest for physics.
The present paper is interested in the micro-distribution of the BEC with spin degrees of freedom and the SSB that is not temeprature-induced but driven by a coherent beam of atoms. In the process, the entropy of the BEC can be transferred to the current of atoms through the quantum entanglement between the propagating atoms and the BEC when they interact. We demonstrate that: (i) a symmetric incident beam can drive the BEC to the SSB phase and the entropy production rate attains its maximum at the instance when the magnetization as the indication of broken spin up-down symmetry becomes nonzero; (ii) the steady spin distributions of the BEC determined by the coherent states of incident beam can be different from the microcanonical one.

The BECs composed of atoms with spin and referred to as spinor BECs (SBECs) possesses large internal degrees of freedom.\cite{stam98,sten98,hall98a,hall98b} Spin-mixing experiments have confirmed that the spin states of SBECs could remain coherent for seconds.\cite{schm04,chan05,law98,pu99} The SBEC degenerated in magnetic polarizations has finite entropy even when its spatial mode is frozen. A rich variety of phases that relay on inter-particle interactions and interactions with external fields have been predicted.\cite{ho98,ohmi98,ueda1,ueda2} Properties of the ground state can be changed dramatically by the Feshbach resonance.\cite{haml09,ho04,land13,hori17,sant17}  Near the resonance, the magnitude of scattering length of two atoms diverges. Additionally, it is enhanced by a factor proportional to the number of atoms due to their coherence in the condensate.\cite{li18} Since the condensate may consist of millions of atoms, the atom-condensate scattering should have a non-negligible probability and thus would be experimentally feasible.

We assume that the spin state of the incoming atoms is a pure quantum state, hence the input entropy is zero. On the other hand, the outgoing atom is entangled with the condensate, forming a finite entanglement entropy. Therefore the atom beam contributes a negative entropy current to the condensate as long as its state is time-dependent.
We further assume that the incident atoms have very low kinetic energy and will thus neither excite spatial modes of the condensate nor lead to heating and/or atom loss of the condensate. Thereby, the entropy associated with the entanglement of the condensate and scattered atom  originates from spin exchange which in turn relies crucially on the principle of identical particles in quantum mechanics and internal symmetries of the system. This entropy appears even if the spin-dependent interaction is negligible during the scattering process.

Two major mechanisms for entropy production in the condensate involve two characteristic times: the scattering time of a propagating atom $t_{s}$ and the decoherent time $t_{d}$.  In the scattering, the outgoing atom shares some information of the condensate and entangles with the latter. The entanglement entropy $S_{e}$ is produced at the same time. On the other hand, there exists an unavoidable weak interaction between the condensate and environment, for instance through the residual magnetic field or an instability of the optical trap that could cause decoherence of the condensate. The time $t_{d}$ is the mean time that the decoherence would happen, accomplishing the thermal entropy $S_{th}$. We will suppose that $t_{d}<t_s$. This means that the condensate is always decoherent after each scattering.

The SBEC consisting of $N$ atoms, each having spin $f=1$, is confined in an optical trap. We adopt the single mode approximation.\cite{law98,gold98,koas00,ho00,duan02} In the absence of an external magnetic field, the energies are proportional to $s(s+1)$ with $s$ being the spin of the condensate. We further assume that the lowest energy level has maximum spin $s=N$ and is degenerate in magnetic polarizations. This is the case of the dilute atom gas with ferromagnetic spin-spin interaction, such as that of $^{87}$Rb. In principle the spin of the condensate may change from $s=N$ to $N-2$ in the scattering. However, the scattering branch ratios of channels between $s=N-2$ and $s=N$ are suppressed by a factor $\frac{1}{N}$ and therefore these channels can be neglected when $N$ is large.\cite{li18} In other words, the energy of the condensate does not change in the scattering.

Since the coupling strength of spin-independent interaction is much larger than that of the spin-dependent interaction,\cite{ueda1,ueda2} we neglect the latter. Denote by $a^{+}_{\mu}$ and $a_{\mu}$ ($c^{+}_{\mu}$ and $c_{\mu}$) creation and annihilation operators of atom in the propagating mode (condensation mode) with polarization $\mu$. The Hamiltonian related to the scattering is given by
\begin{equation}
H_{i}=g_{0}\sum_{\mu_{1}\mu_{2}}a^{+}_{\mu_{1}}c^{+}_{\mu_{2}}(a_{\mu_{1}}c_{\mu_{2}}+a_{\mu_{2}}c_{\mu_{1}})
\label{H}
\end{equation}
The spin-independent coupling $g_{0}$, which has absorbed the integral of the spatial modes, could be varied by the technique of Feshbach resonance. The second term in the bracket of (\ref{H}) responsible for the spin-flip is originated from the effect of identical particles.

The scattering amplitudes can be calculated in the Born approximation as the interaction is short ranged. The spin degrees of freedom, however, need to be treated rigorously because the spin symmetry is crucial in our problem. For this reason, we apply the method of fractional parentage coefficients enabling us to go beyond the mean field approximation.\cite{bao04,li18}

In the proposed experiment all incoming atoms come from the same pure state $|\xi^{i}\rangle_{a}=\sum_{\nu=-1}^{1}\xi^{i}_{\nu}a^{+}_{\nu}|0\rangle_{a}$,
with the subscript $a$ labelling the spin states of the propagating atom.

The state of the condensate scattered from an arbitrary spin distribution can be derived using the scattering from base-state vectors of the condensate. A base-state vector of the condensate may be chosen as the eigenstate of the $Z$-component of the spin, $|m\rangle_{c}$, which is uniquely specified by the magnetization quantum number $m$ ranging from $-N$ to $N$. The composite system initially in $|m_{i}\rangle_{c}|\xi^{i}\rangle_{a}$ will be scattered to
\begin{equation}
|\Psi^{m_{i}}_{\xi^{i}}\rangle=\sum_{m=-N}^{N}\sum_{\mu=-1}^{1}T^{m_{i}}_{m,\mu}|m\rangle_{c}|\mu\rangle_{a}
\label{gsscatted}
\end{equation}
The transition matrix $T^{m_{i}}_{m,\mu}$ is given by
\begin{equation}
T^{m_{i}}_{m,\mu}=(\delta_{m,m_{i}} +C^{N-1,1}_{N,m;m_{i}-\mu,\nu}C^{N-1,1}_{N,m_{i};m_{i}-\mu,\mu})\xi^{i}_{\nu}
\label{Transmatrix}
\end{equation}
Two C-factors being the Clebsch-Gordan coefficients come from the matrix element $\langle m|c^{+}_{\nu}c_{\mu}|m_{i}\rangle$, and
$\nu=m-m_{i}+\mu$ due to the conservation of the total spin polarization.

Generally the condensate is not necessary in a pure state and should be described by a spin state operator. Denote the spin state operator of the condensate after the $n$-th scattering as $\rho_{c}(t_{n})$. Thus the initial state of the condensate prior to the $n$-th scattering is $\rho_{c}(t_{n-1})$. Since $t_{d}<t_s$ has been assumed, $\rho_{c}(t_{n})$ is diagonal in the representation of magnetization. Let its diagonal elements be $\lambda_{m}(t_{n})$.
When the outgoing atom is still entangled with the condensate, i.e., for $t_{n-1}<t<t_{d}$, the scattered atom-condensate system is described by the composite spin state operator $\rho(t)=\sum_{m_{i}}\lambda_{m_{i}}(t_{n-1})|\Psi^{m_{i}}_{\xi^{i}}\rangle \langle\Psi^{m_{i}}_{\xi^{i}}|$. The off-diagonal elements of $\rho(t)$ represent the interference of states with different magnetizations. In this period, the local properties of the condensate are described by the reduced spin operator $\rho^{\prime}_{c}(t)=\mbox{Tr}_{a}\rho(t)$ where the trace is over the spin space of the propagating atom. The state $\rho^{\prime}_{c}(t)$ would be partially coherent.
After $t_{d}$, but before the subsequent scattering, a tiny residue external field will destroy the interference and thus the spin state of condensate will be described by $\rho_{c}(t_{n})$, which is $\rho_{c}^{\prime}(t)$ but with all off-diagonal elements equal to zero.

Using (\ref{gsscatted}) one can derive the iteration equation
\begin{equation}
\lambda_{m}(t_{n})=\frac{1}{Z(t_{n})} \sum_{m^{\prime}}\sum_{\mu=-1}^{1}
|T^{m^{\prime}}_{m,\mu}|^{2}\lambda_{m^{\prime}}(t_{n-1})
\label{rmcond3}
\end{equation}
where $Z(t_{n})$ preserves normalization $\sum_{m}\lambda_{m}(t_{n})=1$.

The reduced state operator of outgoing atom $\rho_{a}(t_{n})$ is derived by tracing out the spin states of the condensate
\begin{equation}
\rho_{a}(t_{n}) =\frac{1}{Z(t_{n})}\sum_{m_{i}}T^{m_{i}}_{m,\mu}T^{m_{i}*}_{m,\mu^{\prime}}\lambda_{m_{i}}(t_{n-1})|\mu\rangle_{a}~_{a} \langle \mu^{\prime}|
\label{rmatom}
\end{equation}
where summations over double repeated indexes $m,\mu,\mu^{\prime}$ are implied. The time-dependent entanglement entropy can be obtained through $\rho_{a}(t_{n})$ as
$S_{e}(t_{n})=-\mbox{Tr}\left[\rho_{a}(t_{n}) \ln \rho_{a}(t_{n})\right]
$.
It is transferred to the environment when the outgoing atom is ``detected" by the environment.
On the other hand, the decoherence appears within $t_{d}$ and leads to the thermal entropy,
$S_{th}(t_{n})=-\sum_{m}\lambda_{m}(t_{n})\ln\lambda_{m}(t_{n})$.
The entropy production rate is the total entropy produced in the $n$-th scattering,
$S_{p}(t_{n})=S_{e}(t_{n})+S_{th}(t_{n})-S_{th}(t_{n-1})
$.
After a sufficiently long time the condensate is driven
to a steady state characterized by a constant thermal entropy whereas the entropy production rate converges to a constant entanglement entropy.

\begin{figure}[tbp]
\includegraphics[width=3 in]{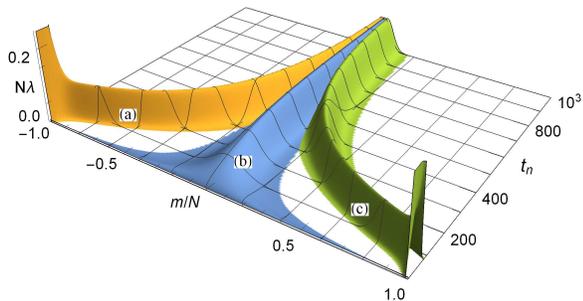}
\caption{The time-dependent spin distributions of spin-1 condensate of $N=100$ spinors from initial distributions: (a) $\lambda_{m}(0)=\delta_{m,-N}$, (b) $\lambda_{m}(0)=\frac{1}{2N+1}$, and (c) $\lambda_{m}(0)=\delta_{m,N}$, respectively. The condensate is scattered by the same species of atoms with incoming state $|0\rangle_{a}$. The axes $\frac{m}{N}$ is the magnetization quantum number per atom. The time $t_{n}=n$ is in the unit of times of scattering.}
\label{fig1}
\end{figure}

Let us first specify the incoming state to $|\xi^{0}\rangle_{a}=|0\rangle_{a}$ and compare the evolutions of three typical initial states of the condensate: (a) $\lambda_{m}(0)=\delta_{m,-N}$ (the negative fully-ordered distribution), (b) $\lambda_{m}(0)=\frac{1}{2N+1}$ (the microcanonical distribution), and (c) $\lambda_{m}(0)=\delta_{m,N}$ (the positive fully-ordered distribution). The time-dependent spin distributions of $N=100$ from the three initial states are plotted in FIG. \ref{fig1}.

\begin{figure}[tbp]
\includegraphics[width=3 in]{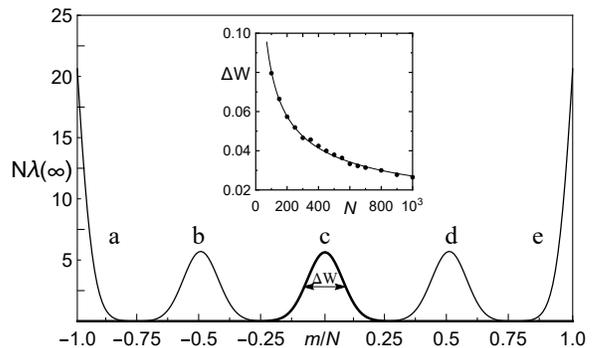}
\caption{The steady spin distributions for spin-1 condensate of $N\leq 10^3$ spinors under the scattering of the atom beam with incoming states $\frac{1}{\sqrt{2}}(|0\rangle_{a}+\cos\theta|-1\rangle_{a}+\sin\theta|1\rangle_{a})$. Curves: (a), (b), (c), (d) and (e) correspond to $\theta=0,\frac{\pi}{6},\frac{\pi}{4},\frac{\pi}{3}, \frac{\pi}{2}$, respectively.  The inset is the half-width of the distribution (c).}
\label{fig2}
\end{figure}

After a sufficiently long time, all distributions converge to steady distributions that are independent of the initial spin state of the condensate but completely determined by the incoming atom states. The steady spin distributions for incoming states $|\xi\rangle_{a}=\sum_{\mu=-1}^1\alpha_{\mu}|\mu\rangle_{a}$ are presented in FIG. \ref{fig2}. It is clearly seen that these distributions are neither microcanonical nor canonical. The steady spin distribution for $|0\rangle_{a}$ and $\frac{1}{\sqrt{2}}|0\rangle_{a}+\frac{1}{2}(|-1\rangle_{a}+|1\rangle_{a})$ appear to be Gaussian-like with the half-width approximately proportional to $\frac{1}{\sqrt{N}}$, see the inset of FIG. \ref{fig2}.

\begin{figure}[tbp]
\includegraphics[width=2.8in]{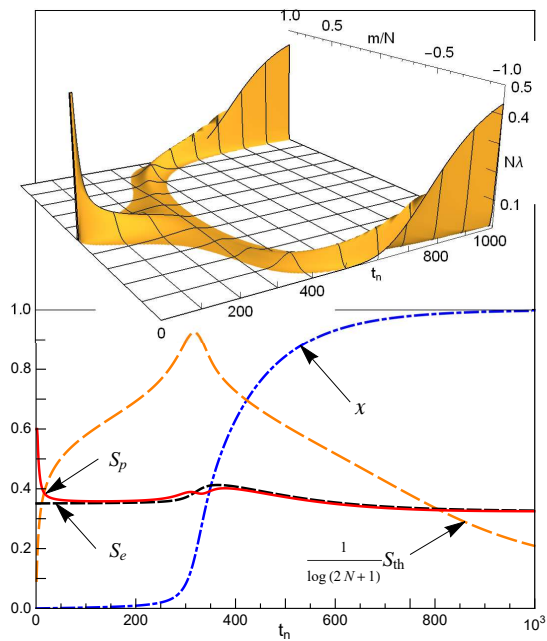}
\caption{The spin-1 condensate of $N=100$ is scattered from the non-polarized initial spin distribution $\lambda_{m}(0)=\delta_{m,0}$ by the same species of atoms in the incoming state $|\xi^{0y}\rangle_{a}=\frac{1}{\sqrt{2}}(|-1\rangle_{a}+|1\rangle_{a})$. In the lower panel, the curves are the time-dependent magnetization fluctuation (blue dash-dotted), the thermal entropy divided by $\log(2N+1)$  (orange long-dashed), the entanglement entropy (red solid), and the entropy production rate (black short-dashed). The upper panel is the time-dependent spin distribution. }
\label{fig3}
\end{figure}

The system has a spin-reversal symmetry and we will show that this symmetry can be broken in course of the evolution. Consider the condensate having disordered initial spin distribution $\lambda_{m}(0)=\delta_{m,0}$ and the incoming-atom state $|\xi^{0y}|\rangle_{a}=\frac{1}{\sqrt{2}}(|-1\rangle_{a}+|1\rangle_{a})$. The  state $|\xi^{0y}|\rangle_{a}$ is an eigenstate of the $Y$-component of the atom spin with vanishing projection on the $Y$-direction. The composite system still has the spin-reversal symmetry. The evolving distribution of $\{\lambda_m(t_{n})\}$ is shown as the upper panel of FIG. \ref{fig3}. A bifurcation emerges suddenly at a time of the evolution, indicating the spin-reversal symmetry breaking.
A hint to the symmetry breaking is given by the time-dependent fluctuation of magnetization quantum number $m$ per atom, $\chi(t)=\sum_{m}(\frac{m}{N})^{2} \lambda_{m}(t)$, presented in the lower panel of FIG. \ref{fig3} as the blue-dash-dotted line.
At a macroscopically short time $t_{0}=\tau N$, with a finite $\tau\ll 1$, the process resembles a random walk on $\left\{x_m=\frac{m}{N}: m=-N,\dots,N\right\}$, therefore the spin distribution is a Gaussian-like centered at zero and cumulants of $\frac{m}{N}$ of order higher than two vanish. In the large $N$ limit we obtain $\chi(t)\sim \chi(t_{0})[1-\frac{6}{7}\chi(t_{0})(t-t_{0})]^{-1}$ for $t>t_{0}$.\cite{supp1}
Since $\chi(t_{0})=\frac{2D t_{0}}{N^{2}}=\frac{2D\tau}{N}$, with $D$ the random-walk diffusion coefficient, then $\chi(t)\sim \frac{2D\tau}{N} [1-\frac{12D\tau}{7N}(t-t_{0})]^{-1}$  is tiny for a large $N$ but it blows up at $t\sim N$. The diverging fluctuation invalidates the Gaussian-like assumption and triggers the bifurcation.

The time-dependent thermal entropy (orange long-dashed line, with the magnitude scaled by $[\log(2N+1)]^{-1}$, the entanglement entropy(red solid line), and the total entropy production rate(black short-dashed line) are also plotted in FIG. \ref{fig3}. Near the bifurcation time, $\chi(t)$ grows rapidly and $S_{th}$ as well as $S_e$ attain maxima.
After a sufficiently long time the condensate will choose one of two fully-ordered states by incident, hence the spin-reversal symmetry is broken. None of the entropies is a monotonic function of time. In the steady state $S_{th}$ does not change and hence $S_p=S_e$.

Summary and discussion: We have investigated the evolution of the spinor Bose-Einstein condensate that is scattered by atoms from a coherent atom source. The dynamics is solely governed by the symmetry of the internal degrees of freedom. We obtained the time-dependent: spin distribution of the condensate, magnetization, magnetization fluctuation, thermal and entanglement entropies, and total entropy production rate. It is found that the steady states of the condensate do not depend on the initial state of the condensate but are determined by the incoming state of the incident atoms. This provides a possible way to prepare various spin distributions of the condensate by choosing suitable incoming states. A novel spontaneous symmetry breaking phenomenon is found in the condensate evolution with the initial magnetization quantum number being zero and the incoming atom state being a symmetric coherent superposition of $|-1\rangle_{a}$ and $|1\rangle_{a}$. The symmetry breaking accompanied with the bifurcation of spin distribution emerges at a mesoscopic time of about twice the atom number of the condensate. After that time the spin distribution develops into two ordered states that have opposite magnetizations. The results would add new insight to spontaneous symmetry breaking and entropy production in small quantum systems. To derive the above results we have assumed that the temperature of the condensate is lower than the excitation energy and the incident atoms have very small kinetic energy so that the condensate remains in the degenerate space of ground states. The required sub-nano Kelvin temperature would be realized in the future. The other less strict assumption is that prior to each scattering the condensate is decoherent, i.e., $t_{d}<t_{s}$. Our study can be generalized to the reverse case of $t_{d}>t_{s}$.

\begin{acknowledgments}
Z.D. thanks the hospitality of the Sun Yat-sen University. The project is supported by the National Key Research and Development Project of China
(Grant: 2016YFA0202001)
\end{acknowledgments}

\clearpage

\end{document}